\documentstyle[12pt,epsfig]{article}
\newcommand{\institute}[1]{\parbox{16cm}{%
\centering\normalsize \sl #1}}

\title{
\bf The strength of the electroweak  \\
phase transition at $m_H \approx$ 80 GeV}
\author{%
F.~Csikor\\\\
\institute{Institute for Theoretical Physics, E\"otv\"os University,\\
H-1088 Budapest, Hungary}\\\\
Z.~Fodor%
\thanks{On leave from Institute for Theoretical Physics,
E\"otv\"os University, H-1088 Budapest, Hungary}\\\\
\institute{KEK, Theory Group, 1-1 Oho, Tsukuba 305, Japan}\\\\
J.~Heitger%
\thanks{Present address: DESY, Platanenallee 6, D-15738 Zeuthen, Germany}\\\\
\institute{Institut f\"ur Theoretische Physik I, Universit\"at M\"unster,\\
D-48149 M\"unster, Germany
}}
\date{}
\begin{document}
\maketitle


\begin{abstract} 
\noindent
In this letter we present the results of our numerical simulations
for the finite temperature electroweak phase transition using the 
SU(2)-Higgs model on four-dimensional lattices at $m_H \approx 80$ GeV.
The temporal extension $L_t=2$ is used for asymmetric lattice spacings 
with an asymmetry parameter $a_s/a_t \approx 4$. The measured 
thermodynamical quantities (interface tension, jump of the order parameter
and latent heat) suggest that the phase transition is of very weakly first 
order. 

PACS Numbers: 11.15.Ha, 12.15.-y\\
\end{abstract}

\vfill

\newpage
\section{Introduction}
The standard picture of the electroweak
theory tells us that at high temperatures (in the early universe)
the electroweak symmetry is restored.  As the universe expands
and cools down, there is a phase transition between the high temperature
``symmetric'' and the low temperature ``broken'' phases. The  
rate of the baryon violating sphaleron processes is unsuppressed at 
high temperatures and is basically  
frozen in the low temperature phase. As a consequence, the presently
observed cosmological baryon asymmetry has been finally determined
at the  electroweak phase transition
\cite{kuzmin}.

In order to clarify 
the details of this phase transition 
several techniques have been used.
Since the bosonic fields have bad infrared behaviour in the perturbative 
approach \cite{perturb,BFHW94,FH94}, several nonperturbative approaches 
have been applied to solve
the problem. The most promising one is the use of lattice simulations.
 Two main 
strategies are used to analyze the problem on the lattice.

One of them is to use four-dimensional lattices
and study the phase transition and its thermodynamical properties there.
Since the bad infrared behaviour is connected to the bosonic
fields only, the SU(2)-Higgs model is analyzed \cite{4d,4d-sigma}, and the 
fermionic sector is included by perturbative steps. It has 
been shown that the finite temperature electroweak phase
transition is of first order for Higgs boson masses around and
below 50 GeV. The strength of the phase transition rapidly
decreases as $m_H$ increases. 

The other possibility contains a systematic combination of
perturbative and non-perturbative methods. 
One starts with the original theory (e.g. the Standard Model)
and integrates out the heavy degrees of freedom perturbatively.
The obtained theory is a three-dimensional bosonic one
(e.g. SU(2)-Higgs or SU(2)$\times$U(1)-Higgs)   \cite{3d}. 

Analyzing the finite temperature electroweak phase transition
by these two different approaches, thus in four and three dimensions,
provides not only a useful cross-check between them, but the comparison of
the results also reveals a lot about the applicability of the perturbative
reduction techniques. Since none of the lattice results contains the 
fermions, the fermionic sector has to be included perturbatively.
This would be completely analogous to the perturbative 
dimensional reduction step for the heavy bosonic modes. 

There has been a lot of speculations that the first order
nature of the electroweak phase transition disappears
for large Higgs boson masses (about 80-90 GeV).
According to $L_t=2$, four-dimensional results with finite-size
analysis there is a first order phase transition for Higgs
boson masses of 66 GeV, but no sign of a 
first order phase transition could be observed for
 masses larger than 85 GeV (last paper in \cite{4d}).  
Recently, in the three-dimensional theory the endpoint of 
the phase transition has been determined \cite{3d-end}. 
Using the same gauge coupling as in the four-dimensional 
approaches, no first order phase transition exists
for Higgs boson masses above $\sim$65 GeV.

It is of essential importance to clarify the relationship between
the full four-dimensional results and the
reduced three-dimensional ones. For this purpose the analysis of 
the order of the phase transition between 65 GeV and 85 GeV
Higgs boson masses in the former case is a particularly sensitive
possibility.
The expected shift of the endpoint compared to three dimensions may give
informations about the overall uncertainties of the reduction procedure.  

In this letter we present the results of our numerical simulations
for the finite temperature electroweak phase transition using the
SU(2)-Higgs model on four-dimensional lattices at $m_H \approx 80$ GeV.
The temporal extension $L_t=2$ is used for asymmetric lattice spacings
with an asymmetry parameter $a_s/a_t \approx 4$. The measured
thermodynamical quantities (interface tension, jump of the order parameter
and latent heat) suggest that the phase transition is of very weakly first
order. Section 2 gives our definitions and some details
of the simulations. Section 3 contains the analysis of the 
thermodynamical quantities, while Section 4  presents an estimate of 
the endpoint for
the finite temperature electroweak phase transition. In Section 5
we summarize our results and give an outlook.

\section{Lattice formulation and simulation}

We will study the four-dimensional SU(2)-Higgs lattice model for
$L_t=2$ temporal extensions. Since
the phase transition is supposed to be quite weak, we need large
physical volumes in the simulations.
The temporal lattice spacing in physical units
is set by the critical temperature of the phase transition,
$a_t=1/(T_cL_t)$. A large physical volume is ensured by a different
choice for the spatial lattice spacing $a_s > a_t$. 
The asymmetric lattice spacing scenario can be induced by choosing 
different coupling strengths in the action for time-like and space-like 
directions. The action reads
\begin{eqnarray}\label{lattice_action}
S[U,\varphi] &=& \beta_s \sum_{sp}
\left( 1 - {1 \over 2} {\rm Tr\,} U_{sp} \right)
+\beta_t \sum_{tp}
\left( 1 - {1 \over 2} {\rm Tr\,} U_{tp} \right)
\nonumber \\
&&+ \sum_x \left\{ {1 \over 2}{\rm Tr\,}(\varphi_x^+\varphi_x)+
\lambda \left[ {1 \over 2}{\rm Tr\,}(\varphi_x^+\varphi_x) - 1 \right]^2
\right. \nonumber \\
&&\left.
-\kappa_s\sum_{\mu=1}^3
{\rm Tr\,}(\varphi^+_{x+\hat{\mu}}U_{x,\mu}\,\varphi_x)
-\kappa_t {\rm Tr\,}(\varphi^+_{x+\hat{4}}U_{x,4}\,\varphi_x)\right\},
\end{eqnarray}
where $U_{x,\mu}$ denotes the SU(2) gauge link variable,  $U_{sp}$ and
$U_{tp}$
the path-ordered product of the four $U_{x,\mu}$ around a
space-space or space-time plaquette, respectively;
$\varphi_x$ stands for the Higgs field. It is
useful to introduce the hopping parameter
$\kappa^2=\kappa_s\kappa_t$ and
$\beta^2=\beta_s\beta_t$. The anisotropies
$\gamma_\beta^2=\beta_t/\beta_s$ and $\gamma_\kappa^2=\kappa_t/\kappa_s$
are functions of the asymmetry $\xi$. These functions have been
determined with identical results on the one-loop level 
perturbatively \cite{T0pert} and non-perturbatively
\cite{T0nonpert} demanding the restoration of the rotational
symmetry in different physical channels. 
 In this paper we use the
specific asymmetry parameter
$\xi=4.052$, which gives $\gamma_\kappa=4$ and $\gamma_\beta=3.919$.
Details of the simulation techniques can be found in \cite{4d}.

We choose  the  parameters as follows. The gauge
coupling is set close to its physical value via $\beta=8$. As it will
be discussed later, its renormalized value receives about 15\% correction.
The scalar self-coupling parameter was chosen to be 
$\lambda=1.92\cdot 10^{-4}$.
For the determination of the critical hopping parameter $\kappa_c$ 
and the interface tension (see next section) we have used the 
two-coupling method.
The final value for the transition point from this
procedure on a $2\times24^2\times192$ lattice gives  
$\kappa_c=0.107791(3)$.
To study final volume behaviour we also simulated on a larger lattice, namely 
$2\times32^2\times288$, where we obtained $\kappa_c=0.1077835(25)$.

In order to fix the physical parameters in a numerical simulation,
one has to define and compute  suitable renormalized quantities
at zero temperature.
The renormalized gauge coupling can be determined from the static
potential of an external SU(2) charge pair, measured by Wilson loops.
The physical Higgs mass $m_H$ can be extracted from correlation
functions of e.g.  $R_x \equiv {1 \over 2}{\rm Tr\,}(\varphi_x^+\varphi_x)$ and 
the  $\varphi$--link operators 
$ L_{\varphi,x} \equiv
\kappa_s\sum_{\mu=1}^3
{\rm Tr\,}(\varphi^+_{x+\hat{\mu}}U_{x,\mu}\,\varphi_x)
+\kappa_t {\rm Tr\,}(\varphi^+_{x+\hat{4}}U_{x,4}\,\varphi_x)\,.$
The W-boson mass $m_W$ can be obtained similarly from the composite
link fields ($\varphi_x=\rho_x \cdot \alpha_x$ where $\alpha_x$ is an
SU(2) matrix and $\rho_x\ge 0$)
$W_{x;rk}\equiv
\frac{1}{2}\,{\rm Tr\,}
\left(\tau_r\alpha^+_{x+\hat{k}}U_{x,k}\,\alpha_x\right)
\,,\quad
\mbox{$\tau_r$: Pauli matrices}\,,\quad
r,k=1,2,3\,.$
The connected correlation functions of these operators have
been measured in direction of largest extension on  two lattices. 
Note that in principle all numerical estimates for the lightest states in 
a given channel are only upper bounds. Since at our parameters we are 
deep in the Higgs phase when simulating the $T=0$ quantities, the lowest 
mass is well separated from the vacuum and a cross correlation matrix 
analysis does not seem to be very essential. 
The mass fitting was done by the Michael-McKerrel method 
\cite{M94MMc95}, whose
features and application in the SU(2)-Higgs model have been sketched in
the fourth reference of \cite{4d}. The details of the correlation function
and Wilson loop analysis for anisotropic lattice actions can be found
in \cite{T0nonpert}.

At $T=0$ and $\kappa=\kappa_c$ we obtained the following numerical results
on $16\times8^2\times64$ lattices 
(in units of $a_t$):
the Higgs boson mass is $m_H=0.2711(51)$ and the W-mass is
$m_W=0.2803(39)$. 
Combining them with additional data from simulations in the vicinity
of $\kappa_c$ in order to account for its uncertainty as well, one gets
$R_{HW}\equiv m_H/m_W=0.975(50)$, which corresponds to $m_H=78(4)$ GeV
pole mass, if $m_W=80$ GeV sets the physical scale.
The renormalized gauge coupling comes out to be $g_R^2=0.539(16)$.
Previous experience \cite{4d} shows that the masses and coupling we obtained 
on the above lattice basically coincide with the infinite volume values 
within our errors.

\section{Thermodynamical quantities and results}

Near the endpoint  the electroweak phase transition falls into 
the three dimensional Ising universality class \cite{ujlaine}, 
where the two peak structure persists at any finite volume 
(even at the endpoint). Therefore, we have taken particular attention to study 
finite volume effects and extrapolate to the infinite volume limit.
In view of the large lattices one is confronted with for an extremely
weak first order phase transition also in the case of anisotropic
lattices, the interface tension $\sigma$ has been calculated by
employing the two-coupling method \cite{PR89HPRS9091} in $\kappa$.
This  turned out to be robust and most economic  in our previous analyses as 
compared to the histogram or the tunneling correlation
length methods \cite{4d-sigma}.
Following refs.~\cite{4d,4d-sigma}, the generalization to a situation
with anisotropic lattices is straightforward. Namely, if an interface pair
perpendicular to the $z$--direction is enforced by dividing the lattice
volume in symmetric and Higgs phases with
\begin{equation}
\kappa=(\kappa_1,\kappa_2)=
(\kappa_1<\kappa_c: z\le L_t/2\,,\,\kappa_2>\kappa_c: z>L_t/2)
\end{equation}
as for the $\kappa_c$--determination, the related additional free
energy $\Delta F$ yields for $\Delta\kappa\equiv|\kappa_i-\kappa_c|\ll1$
the expression
\begin{equation}\label{sigma}
a_s^2a_t\sigma=
\lim_{\Delta \kappa \rightarrow 0}
\left\{\left(\Delta \kappa\right)
\lim_{L_z\to\infty}L_z\cdot\Delta L_\varphi(\kappa_1,\kappa_2)\right\}\,,
\end{equation}
where $L_{\varphi}^{(i)}=L_{\varphi}^{(i)}(\kappa_1,\kappa_2)$, $i=1,2$,
denote the expectation value of the $\varphi$--link operators
 in the respective phases and
$\Delta L_{\varphi}(\kappa_1,\kappa_2)\equiv
L^{(2)}_{\varphi}(\kappa_1,\kappa_2)
-L^{(1)}_{\varphi}(\kappa_1,\kappa_2)$ their difference.
Now, since the free energy shift can be shown to behave as
$\Delta F\approx{\cal O}(\Delta\kappa)$, the $(N+2)$--parametric
Laurent ans\"atze
\begin{equation}\label{laurent}
L_{\varphi}^{(i)}(\kappa_1,\kappa_2)=
-\frac{c_i}{\kappa_i-\kappa_c}
+\sum_{j=0}^{N}\gamma^{(j)}_i(\kappa_i-\kappa_c)^j
+{\cal O}\left((\Delta\kappa)^{N+1}\right)\,,\quad i=1,2
\end{equation}
give the finite volume estimator for the interface tension
$a_s^2a_t\hat{\sigma}=L_z\,(c_1+c_2)\,.$  

\begin{figure}[htb]
\begin{center}
\epsfig{file=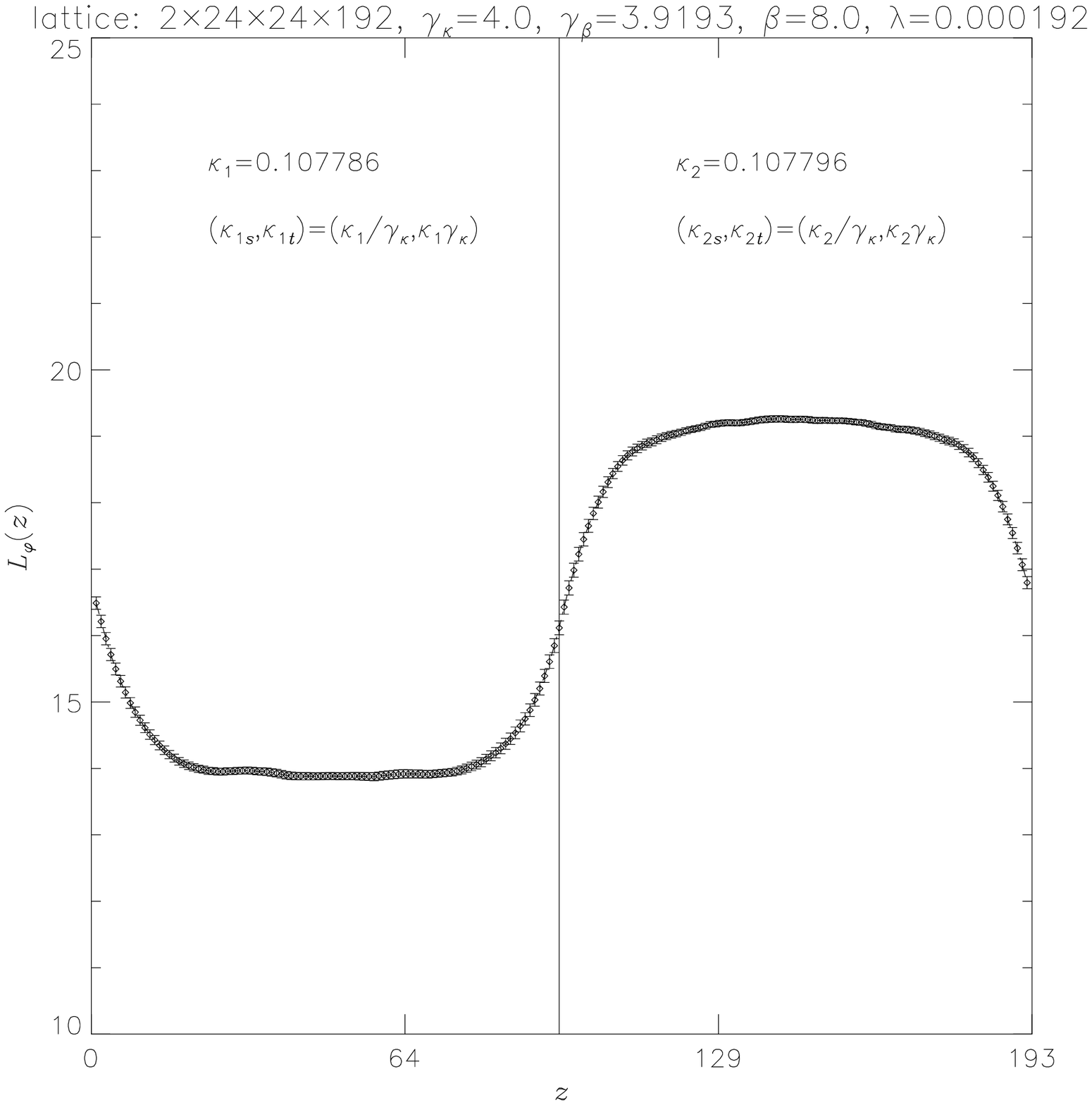,width=8cm,height=9.525cm}
\epsfig{file=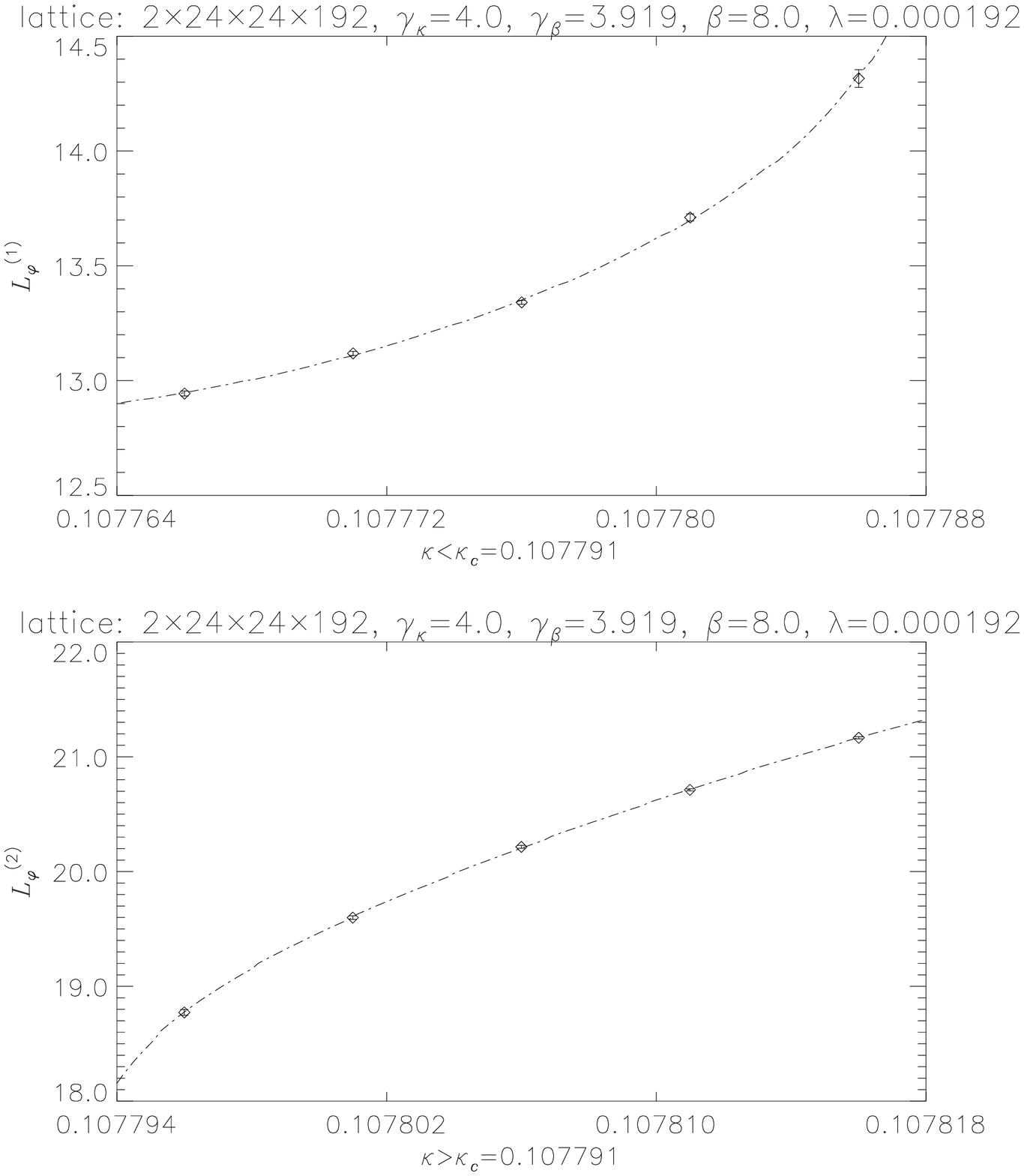,width=8cm}
\parbox{\textwidth}{
\caption{\label{LphiPlot} Left:
                          two-phase profile of the $z$--slice
                          expectation value $L_{\varphi}(z)$ of
                          $L_{\varphi}$ 
                          for the smallest $\kappa$--interval used in the
                          simulations for the $\sigma$--determination in
                          table~\ref{ResLphi}.
                          Right:
                          four-parameter least-squares fits of
                          $L_{\varphi}^{(i)}$, $i=1,2$, separately in each
                          phase. The $\chi^2$--values are $\chi_1^2=2.59$
                          and $\chi_2^2=0.96$, respectively.}
}
\end{center}
\end{figure}

In the left part of figure~\ref{LphiPlot} we illustrate a characteristic
two-phase distribution of $z$--slices for the expectation value
$L_{\varphi}(z)$.
An interface pair has developed, and the plateaus marking the pure-phase
expectation values $L_{\varphi}^{(i)}(\kappa_1,\kappa_2)$ are flat
and broad enough to ensure that the coexisting phases are stable
against any turn-over into one single phase.
All simulation data, which were collected in $\kappa=(\kappa_1,\kappa_2)$
centred symmetrically around $\kappa_c$ for the determination of
$\sigma$, are listed in table~\ref{ResLphi}.
%
\begin{table}[htb]
\begin{center}
\begin{tabular}{|c|c|c||c|c|c|}
\hline
  $\kappa_1$ & $\kappa_2$ & sweeps & $L_{\varphi}^{(1)}$
& $L_{\varphi}^{(2)}$ & $\Delta L_{\varphi}$ \\
\hline\hline
  0.107761 & 0.107821 & 40000 & \it{12.7856(73)} & \it{21.5913(92)} 
& \it{8.8057(90)} \\
  0.107766 & 0.107816 & 40000 & 12.944(11)  & 21.1656(99) & 8.2217(95) \\
  0.107771 & 0.107811 & 60000 & 13.1186(89) & 20.7117(90) & 7.593(11)  \\
  0.107776 & 0.107806 & 60000 & 13.3410(90) & 20.215(13)  & 6.874(12)  \\
  0.107781 & 0.107801 & 70000 & 13.711(15)  & 19.599(16)  & 5.888(20)  \\
  0.107786 & 0.107796 & 80000 & 14.316(39)  & 18.773(25)  & 4.457(37)  \\
\hline
\end{tabular}
\parbox{\textwidth}{
\caption{\label{ResLphi} Results for $L_{\varphi}^{(1)}$,
                         $L_{\varphi}^{(2)}$ and $\Delta L_{\varphi}$
                         for the calculation of $\sigma$ on a lattice of
                         size $2\times24^2\times192$, together with their
                         statistical errors from a binning procedure.
                         In order to obtain acceptable $\chi^2$--values,
                         the entries in italics were only omitted for the
                         four-parameter fits (see text).}
}
\end{center}
\end{table}
%
The integrated autocorrelation times of the $\varphi$--links are 
 18 to 160 sweeps for our largest
and smallest $\kappa$--interval, respectively.
Compared to previous analyses on isotropic lattices for stronger first
order transitions, the $\kappa$--intervals have been chosen closer to
each other and to $\kappa_c$, while it was always verified that the
distinct phases were well separated.
At a very weak transition 
 the diverging of $L_{\varphi}^{(i)}$, when $\kappa_1$ ($\kappa_2$)
approach $\kappa_c$ from below (above), sets in
later and will be less pronounced than for smaller Higgs masses.
Therefore, the curvatures in the shapes of the two $L_{\varphi}^{(i)}$
as functions of $\kappa$ and particularly their residua 
should be modelled more accurately in the present case.
The least-squares fits to eq.~(\ref{laurent}) can be done for both
$\varphi$--links individually or for their difference
$\Delta L_{\varphi}$.
We observed \cite{4d-sigma}  that
the second alternative should be preferred.
Firstly,  correlations
between the competing phases give fluctuations in
the locations of the interfaces and secondly, a roughening of the
interfaces is expected as the phase transition weakens. The two effects are 
supposed to compensate to some extent in $\Delta L_{\varphi}$.

\begin{figure}[htb]
\begin{center}
\epsfig{file=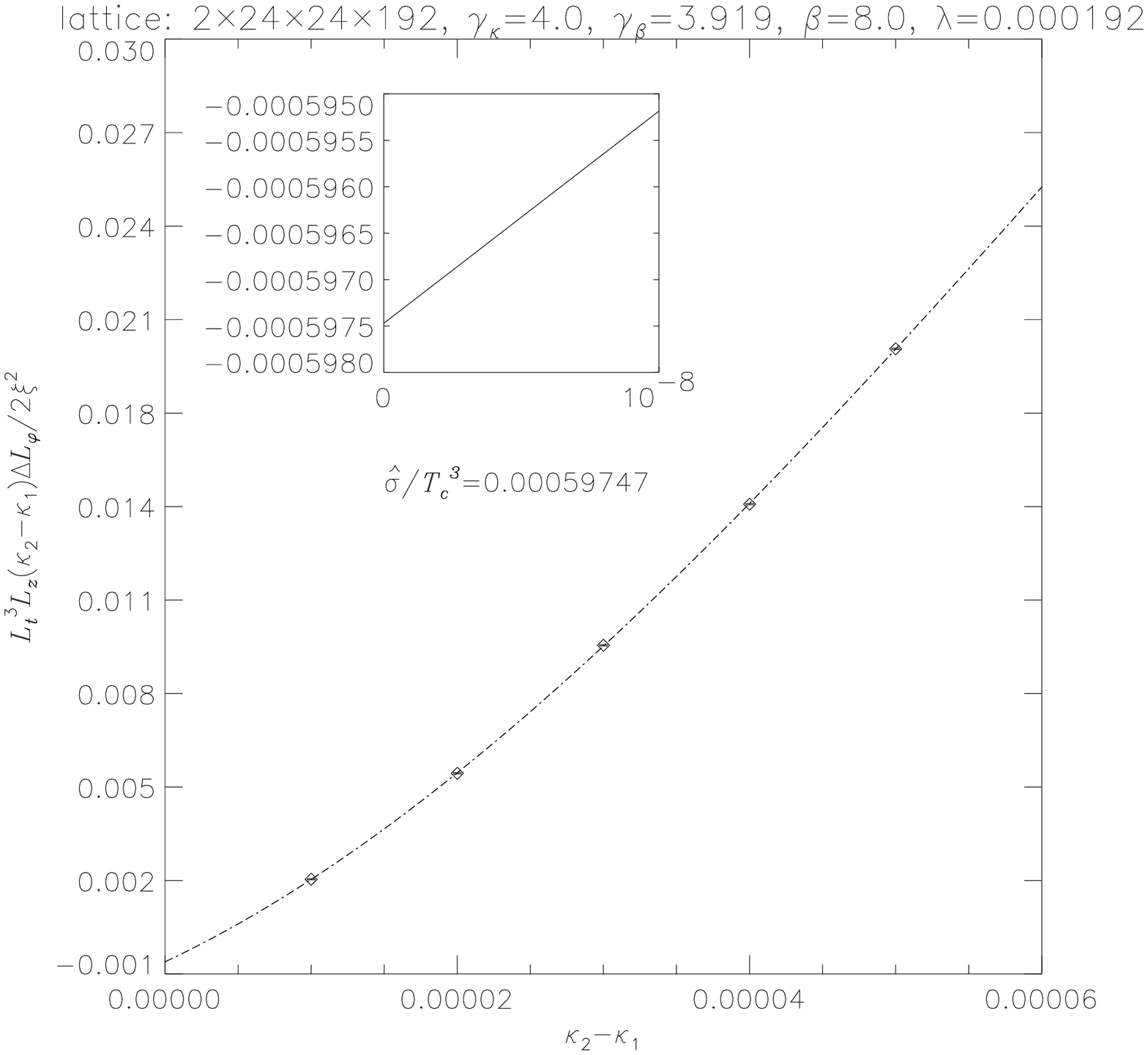,width=12cm}
\parbox{\textwidth}{
\caption{\label{DLphiPlot} Four-parameter least-squares fit of the
                           $\varphi$--link difference $\Delta L_{\varphi}$
                           as a function of $\kappa_2-\kappa_1$ with 
                           $\chi^2=4.55$.
                           The extrapolation to $\kappa_2-\kappa_1=0$
                           gives the interface tension.}
}
\end{center}
\end{figure}

In this spirit we carefully examined nearly all types of fits from
three- to six-parametric in the Laurent ans\"atze of eq.~(\ref{laurent})
for the two kinds of observables $L_{\varphi}^{(i)}$ and
$\Delta L_{\varphi}$ and for every allowed fit intervals, which could
be built up.
In part the results on $\sigma$ might depend on the specific fit
interval and on the number of fit parameters in question, but if
the inverse-linear part of the general ansatz is supplemented with a
fourth or even a fifth fit parameter, the fits are acceptable throughout
with tolerable $\chi^2/\mbox{dof}$ and can readily be extended over
almost the whole range of $\kappa$--pairs.
In order to control the relevance of a further degree of freedom, we
also performed some fits with an additional parameter multiplying the
next higher power in $\kappa_i-\kappa_c$. 
This gave either unreasonably too large $\chi^2/\mbox{dof}$--values,
or the error analysis of these fits led to very high statistical errors
so that the new fit parameters could not be resolved reliably.
As exemplarily displayed in the right diagrams of figure~\ref{LphiPlot}
and in figure~\ref{DLphiPlot}, we found for the best fits
$\left(\hat{\sigma}/T_c^3\right)_{L_{\varphi}}=0.00060(8+22)$ from
four-parameter $\chi^2$--fits of $L_{\varphi}^{(i)}$, $i=1,2$, and
$\left(\hat{\sigma}/T_c^3\right)_{\Delta L_{\varphi}}=0.00060(7)$
from the similar fit of $\Delta L_{\varphi}$.
%
%
%
%
The quoted error on the former incorporates the statistical error
from a bootstrap analysis \cite{E79,4d-sigma} and an error caused
by the uncertainty in $\kappa_c$ (second entry in parentheses),
whereas the solely statistical error on the latter stems from repeated
fits with 1000 normally distributed random data around the measured mean
values as input.

To study  possible finite volume effects and the infinite volume limit, 
we also simulated
 on a larger lattice of size $2\times 32^2\times 288$ at otherwise
unchanged parameters and couplings. The data are displayed in table 2. 
After an analogous evaluation we obtained for 
$\left(\hat{\sigma}/T_c^3\right)_{L_{\varphi}}=0.00062(10+25)$ from
four-parameter $\chi^2$--fits of $L_{\varphi}^{(i)}$, $i=1,2$, and
$\left(\hat{\sigma}/T_c^3\right)_{\Delta L_{\varphi}}=0.00063(8)$
from the similar fit of $\Delta L_{\varphi}$.
These are fully compatible within errors to the preceding results, which
indicates that the surface tension does not decrease 
for increasing volume when the volume is encreased by a factor of 2.7. Thus 
data are compatible with a first order phase transition.

\begin{table}[htb]
\begin{center}
\begin{tabular}{|c|c|c||c|c|c|}
\hline
  $\kappa_1$ & $\kappa_2$ & sweeps & $L_{\varphi}^{(1)}$
& $L_{\varphi}^{(2)}$ & $\Delta L_{\varphi}$ \\
\hline\hline
  0.107757 & 0.107810 & 10000 & 12.6839(58) & 21.146(12) & 8.462(13) \\
  0.107762 & 0.107805 & 10000 & 12.8446(97)  & 20.708(11) & 7.863(13) \\
  0.107767 & 0.107800 & 10000 & 13.026(12) & 20.194(19) & 7.168(20)  \\
  0.107772 & 0.107795 & 10000 & 13.293(17) & 19.592(25)  & 6.299(23)  \\
  0.107777 & 0.107790 & 10000 & 13.620(20)  & 18.752(33)  & 5.132(31)  \\
  0.107780 & 0.107787 & 5000  & 14.075(46)  & 18.107(54)  & 4.032(72)  \\
\hline
\end{tabular}
\parbox{\textwidth}{
\caption{\label{ResLphi1} Results for $L_{\varphi}^{(1)}$,
                         $L_{\varphi}^{(2)}$ and $\Delta L_{\varphi}$
                         for the calculation of $\sigma$ on a lattice of
                         size $2\times32^2\times288$, together with their
                         statistical errors from a binning procedure.
}}
\end{center}
\end{table}
To present our final (infinite volume) value we average all these results 
to a combined
estimate and take their absolute spread including the errors as a
measure for the sum of statistical and systematic uncertainties.
Assuming all reasonable fits from both spatial volumes to contribute,
we finally arrive at
\begin{equation}\label{sigma_final}
\left(\frac{\hat{\sigma}}{T_c^3}\right)
^{\scriptsize \mbox{all}}
_{\scriptsize \mbox{2--$\kappa$}}=0.0006(4)\,.
\end{equation}

\begin{figure}[htb]
\begin{center}
\epsfig{file=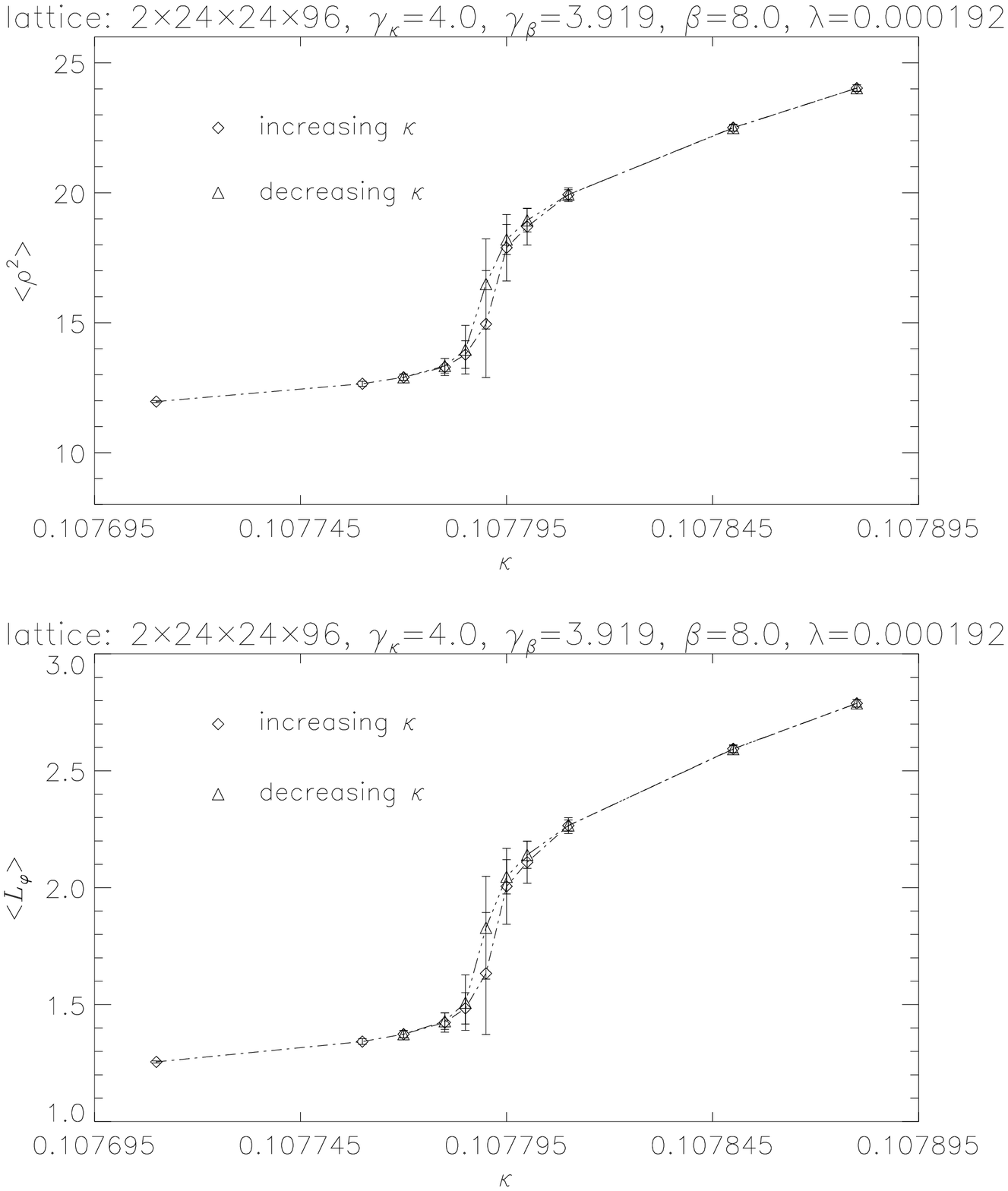,width=12cm}
\parbox{\textwidth}{
\caption{\label{HystPlot} Hysteresis patterns of the operators $\rho^2$
                          and $L_\varphi$ around the critical
                          $\kappa$--region at $L_t=2$.}
}
\end{center}
\end{figure}

From quadratic fits of the discontinuities of the order parameters
showing up in the thermal cycles of figure~\ref{HystPlot}
%
%
for $\rho^2$ and $L_\varphi$ (and similarly for the plaquette variables)
we also extracted the jump in the Higgs field vacuum expectation value  
and the latent heat. The jump of the Higgs field vacuum expectation value 
is given by  $\Delta v/T_c=L_t\xi^{-1}\sqrt{2\kappa\,\Delta\langle\rho^2\rangle}$ in case of asymmetric lattice spacings. The latent heat is determined 
by 
\begin{eqnarray}\label{lheat1}
\frac{\Delta\epsilon}{T_c^4}=  
\frac{L_t^4}{\xi^3}\,\frac{\partial \Delta s}{\partial\tau}\,,
\end{eqnarray}
where $\Delta $ means the jump, $s$ is the average action density and
$\tau\equiv-\ln(a_s m_W)$ and the derivative is taken with constant $\xi$.
To determine the partial derivatives of $\beta_s$, $\beta_t$ and $\lambda$ 
we have 
used the perturbative renormalization group equations defining the 
{\em lines of constant physics},  while for $\kappa$ we have followed a 
procedure similar to the one used in \cite{4d} for the isotropic case.
We end up with
\begin{eqnarray}\label{lheat2}
\frac{\Delta\epsilon}{T_c^4}=
\frac{L_t^4}{\xi^3}\,\left\{\frac{1}{\kappa}
\frac{\partial\kappa}{\partial\tau}\,
\Delta < l_\varphi > \, + \frac{43}{16\pi^2}g^2 (\beta_s 
\Delta <p_s> + \beta_t \Delta <p_t>) \right. \nonumber \\
\left. +\Delta <Q> \left(\frac{2 \lambda}{\kappa} 
\frac{\partial\kappa}{\partial\tau} + \frac{\kappa^2}{4\xi\pi^2}\, 
\Big[96 \lambda_c^2 + \frac{9}{32} g^4-9\lambda_c g^2\Big]\right)\right\},
\end{eqnarray} 
where    
$< l_\varphi >$ is the average link, $<p_s>$ and $<p_t>$ 
are the average plaquette variables, $<Q>$ is the average of
%
%
the quartic part in $\varphi$ of the lattice action
(\ref{lattice_action}) and $\lambda_c=\lambda \xi /(4\kappa^2)$. 
The results for our simulation parameters are collected in
table~\ref{ResTab}.
%
\begin{table}[htb]
\begin{center}
\begin{tabular}{|c||c|c|c|c|}
\hline
  $L_t$ & $T_c/m_H$ & $\hat{\sigma}/T_c^3$ & $\Delta v/T_c$
& $\Delta\epsilon/T_c^4$ \\
\hline\hline
  2 & 1.86(2) & 0.0006(4) & 0.37(16) & 0.0033(27) \\
\hline
\end{tabular}
\parbox{\textwidth}{
\caption{\label{ResTab} Summary of our lattice results at $L_t=2$
                        temporal extension.
}}
\end{center}
\end{table}
%

\section{Estimate for the endpoint for the electroweak phase transition}

It is a well known perturbative feature that the finite
temperature electroweak phase transition is of first order due to the
vanishing magnetic mass ($m_T$). A non-vanishing parametrization of this
mass predicts an endpoint $m_{H,c}$ for the first order phase transition.
Note that our treatment is purely phenomenological and is performed in 
one particular gauge, the Landau gauge.
The value of the magnetic mass can be obtained
e.g. by solving the truncated Dyson-Schwinger equations (gap-equations)    
\cite{BFHW94,BP96}. Different treatments
of these equations give
different numerical values for the magnetic mass. Nevertheless keeping it
as an unknown parameter and analysing the behaviour of the different
static thermodynamical quantities (e.g. jump of the order parameter,
latent heat and interface tension), a scaling behaviour can be observed
in the vicinity of $m_{H,c}$ \cite{BFHW94}. As we have explicitly checked,
this behaviour is essentially independent of the model or loop-order used 
\cite{FH94}.

We have fitted all our available
four-dimensional data at $m_H \approx 19,35,49,80$ GeV (see \cite{4d} and the 
results of the present paper) on the static
thermodynamical quantities to the predictions of the one- and two-loop
finite temperature effective potential. The perturbative 
results were considered as a function of one unknown parameter, namely the
endpoint of the phase transition $m_{H,c}$ parametrized by a
non-vanishing mass in the magnetic sector. 
The small masses play a less relevant role in the fiting procedure, 
while the influence of $49$ and $80$ GeV data on the value 
of the endpoint is more important. Combining all results (i.e.~all
masses and thermodynamical quantities), we have seen that at the optimal
magnetic mass $\chi^2/\mbox{dof}\approx 1$, whereas for vanishing $m_T$
the $\chi^2$--value is considerably larger. This fact can be interpreted as
a sign, which indicates the presence of the phase transition endpoint. 
Our combined value for the endpoint is 
$m_{H,c}=106^{+88}_{-24}$ GeV.  The estimated value for $m_{H,c}$ is
slightly larger than the values of the three-dimensional analyses 
\cite{3d-end};
however, they are in an approximately 1.5--$\sigma$ agreement.
We emphasize that at small $L_t$'s  the phase transition usually tends to be 
stronger than for larger $L_t$ values \cite{4d}.

\section{Discussion and outlook}

Both quantities $\hat{\sigma}/T_c^3$ and $\Delta\epsilon/T_c^4$ for
$m_H\approx 80$ GeV are substantially smaller than the perturbative
predictions with zero magnetic mass (e.g.~$\sigma/T_c^3\approx 0.002$ 
\cite{perturb}).
Comparing with our earlier investigations at lower $m_H$ 
\cite{4d,4d-sigma}, these results confirm the expectation that the interface
tension and the latent heat are steeply decreasing functions of the Higgs
boson mass, and they are even consistent at $m_H\approx 80$ GeV with a no 
first order phase transition scenario approximately at the 1--$\sigma$
level.

We have fitted all our four-dimensional data to the perturbative
predictions assuming, however, an endpoint for the phase transition
parametrized by a phenomenological non-vanishing magnetic mass. 
Our combined value for the endpoint is $m_{H,c}=106^{+88}_{-24}$ GeV.
Although being in a nearly 1.5--$\sigma$ agreement, the fact that our
findings deviate from those of the three-dimensional investigations
\cite{3d-end}, which claim the endpoint of the first order phase transition 
line to be at $m_{H,c}\approx 67$ GeV, should be clarified 
in future.

However,  we also have to concede that a temporal lattice extension of $L_t=2$
may be still too far from the continuum physics.
In fact, experience with lattice perturbation theory shows \cite{T0nonpert} 
that using anisotropic lattices makes the approach to the continuum limit 
even slower than for isotropic lattices.
So at least some knowledge about the behaviour of the four-dimensional
SU(2)--Higgs model at $m_H\approx 80$ and $L_t=3$ is of principal
interest before drawing a final conclusion. Unfortunately the necessary
CPU-time requirements to reach an adequate precision on such huge
lattices seems not yet to be realistic. 
Therefore, it would be more desirable to determine the endpoint of the
electroweak phase transition on the basis   of a study of the Lee-Yang
zeros, as applied in refs.~\cite{3d-end} to the dimensionally reduced
model, within the theory in four dimensions as well.

\section*{Acknowledgements}

Special thanks go to I. Montvay and G. M\"unster for essential proposals.
Discussions with K. Kajantie, F. Karsch, A. Patk\'os, M. Shaposhnikov
and R. Sommer are also acknowledged. 
Simulations have been carried out on the Cray-T90 at HLRZ-J\"ulich,
on the APE-Quadrics at DESY Zeuthen and on the PMS-8G PC-farm in Budapest.
F. Cs. and Z. F. were partially 
supported by Hung. Sci. Foundation Grants T016240/T022929 and FKFP-0128/1997.

\end{document}